\documentclass[twocolumn,showpacs,floatfix,prb,superscriptaddress,eqsecnum]{revtex4}
\usepackage{graphicx}
\usepackage{amsmath}
\usepackage{bm}
\usepackage{psfrag}
\begin{document}

\title{Calculation of the conductance of a graphene sheet using the Chalker-Coddington network model}
\author{I. Snyman}
\affiliation{Instituut-Lorentz, Universiteit Leiden, P.O. Box 9506, 2300 RA Leiden, The Netherlands}
\author{J. Tworzyd{\l}o}
\affiliation{Institute of Theoretical Physics, Warsaw University, Ho{\.z}a 69, 00-681 Warsaw, Poland} 
\author{C. W. J. Beenakker}
\affiliation{Instituut-Lorentz, Universiteit Leiden, P.O. Box 9506, 2300 RA Leiden, The Netherlands}
\date{March 2008}
\begin{abstract}
The Chalker-Coddington network model (introduced originally as a model for percolation in the quantum Hall effect) is known to map onto the
two-dimensional Dirac equation. 
Here we show how the network model can be used to solve a scattering problem in a weakly doped graphene sheet connected
to heavily doped electron reservoirs. 
We develop a numerical procedure to calculate the scattering matrix with the
aide of the network model. For numerical purposes, the advantage of the network model over the honeycomb lattice
is that it eliminates intervalley scattering from the outset.  
We avoid the need to include the heavily doped regions in the network model (which would be computationally
expensive), by means of an analytical relation between the transfer matrix through the weakly doped region and the scattering matrix between the electron
reservoirs.
We test the network algorithm by calculating the conductance of an electrostatically defined quantum point contact and comparing with the
tight-binding model of graphene. We further calculate the conductance of a graphene sheet in
the presence of disorder in the regime where intervalley scattering is suppressed. We find
an increase in conductance that is consistent with previous studies.
Unlike the tight-binding model, the network model does not require smooth potentials 
in order to avoid intervalley scattering.
\end{abstract}
\pacs{73.50.Td, 73.23.-b, 73.23.Ad, 73.63.-b}
\maketitle
\section{Introduction}
\label{sect1} 

The low-energy and long-wave-length properties of conduction electrons
in a carbon monolayer (graphene) are described by the two-dimensional
Dirac equation \cite{Cas07}. In one-dimensional geometries this partial
differential equation can be solved analytically, but fully
two-dimensional problems typically require a discretization to permit a
numerical solution. The tight-binding model on the honeycomb lattice of
carbon atoms provides the most obvious and physically motivated
discretization \cite{McC56}. The band structure of a honeycomb lattice
has two valleys, coupled by potential variations on the scale of the lattice
constant. Smooth potentials are needed if one seeks to avoid
inter-valley scattering and obtain the properties of a single valley.

Discrete representations of the Dirac equation that eliminate from the
outset the coupling to a second valley may provide a more efficient way
to isolate the single-valley properties. Alternative tight-binding
models \cite{Fis85,Lud94,Lee94,Zie95} have been introduced for that
purpose. One method of discretization which has received much attention
is the network model, originally introduced by Chalker and Coddington as
a model for percolation in the quantum Hall effect \cite{Cha88}. Ho and
Chalker \cite{Ho96} showed how a solution of this model can be mapped
onto an eigenstate of the Dirac equation, and this mapping has proven to
be an efficient way to study the localization of Dirac fermions
\cite{Kra05}.

The recently developed capability to do transport measurements in
graphene \cite{Gei07} has renewed the interest in the network model
\cite{Hir07} and also raises some questions which have not been
considered before. The specific issue that we address in this paper is 
how to introduce metallic contacts in the network model of graphene. Metallic
contacts are introduced in the Dirac equation by means of a downward 
potential step of magnitude $U_{\infty}$. The limit $U_{\infty}\rightarrow\infty$ is
taken at the end of the calculation. (It is an essential difference with the 
Schr\"{o}dinger equation that an infinite potential step produces a
\textit{finite} contact resistance in the Dirac equation.) This phenomenological 
model of metallic leads, introduced in Ref.\ \onlinecite{Two06}, is now
commonly used because 1) it is analytically tractable, 2) it introduces no 
free parameter, and 3) it agrees well with more microscopic models
\cite{Sch07,Bla07}. A direct implementation of such a metallic contact in the 
network model is problematic because the mapping onto the Dirac equation
breaks down in the limit $U_{\infty}\rightarrow\infty$. 
Here we show how this difficulty can be circumvented.

To summarize then, there is a need to develop numerical methods for Dirac
fermions in graphene when the potential landscape does not allow analytical
solutions. If one implements a method based on the honeycomb lattice of graphene,
intervalley scattering is present, unless the potential is smooth on the scale of
the lattice. Smooth potential landscapes are experimentally relevant, but computationally
expensive, because they require discretization with a large mesh. It is therefore preferable to
develop a numerical method that eliminates intervalley scattering from the outset.
The known correspondence between the Chalker-Coddington network model and the Dirac equation
provides such a method, as we show in this paper. The key technical result
of our work is an analytical method to include heavily doped reservoirs. (Including
these reservoirs numerically would have been prohibitively expensive, computationally.)  

In Secs.\ \ref{sect2} and \ref{sect3} we summarize the basic equations 
that we will need, first regarding the Dirac equation and then regarding the network
model. Our key technical result in Sec.\ \ref{sect4} is a relationship 
between the scattering problems for the Dirac equation in the limit
$U_{\infty}\rightarrow\infty$ and for the network model at $U_{\infty}\equiv 0$. 
We test the method in Sec.\ \ref{five} by calculating the conductance of
an electrostatically defined constriction (quantum point contact) in a graphene sheet.
We also study the effect of disorder on conductance. We confirm the results of 
previous studies\cite{Ryc07,Bar07,Nom07,San07} that smooth disorder (that does not cause intervalley scattering)
enhances the conductivity of undoped graphene. We conclude in Sec. \ref{conclusion}.

\section{Formulation of the scattering problem}
\label{sect2}
\subsection{Scattering Matrix}
\begin{figure}[t]
  \begin{center}
    \includegraphics[width=.95 \columnwidth]{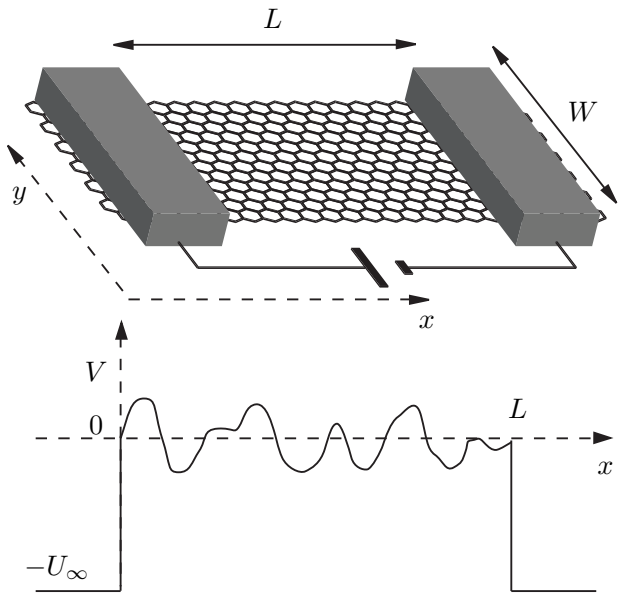}
    \caption{Top panel: Schematic of a graphene sheet contacted by two electrodes. A voltage
source drives a current through the sheet. The bottom panel shows the potential profile $V(x,y)$ for fixed $y$.
\label{figgeom}} 
  \end{center}
\end{figure}
A scattering formulation of electrical conduction through a graphene sheet was given in Ref.~\onlinecite{Two06}.
We summarize the basic equations. The geometry, shown in Fig.~\ref{figgeom}, consists of a
weakly doped graphene sheet (length $L$ and width $W$) connected to heavily doped graphene leads.
A single valley has the Dirac Hamiltonian 
\begin{equation}
H=v\bm{\sigma}\cdot\left[\bm{p}-e\bm{A}(\bm{r})\right]+V(\bm{r})+\sigma_z \mu(\bm{r}),
\end{equation}
where $\bm{A}(\bm r)$ is the magnetic vector potential, $V(\bm{r})$ is the
electrostatic potential, and $\mu(\bm{r})$ is a substrate-induced mass term.
The vector $\bm{\sigma}=(\sigma_x,\sigma_y)$ contains the standard Pauli matrices
\begin{equation}
\sigma_x=\left(\begin{array}{cc}0&1\\1&0\end{array}\right),\hspace{3mm}
\sigma_y=\left(\begin{array}{cc}0&-i\\i&0\end{array}\right).
\end{equation}
We assume that the fields $\bm{A}$, $V$, and $\mu$ are smooth on the scale of the 
lattice constant, so that the valleys are uncoupled.

In the heavily doped leads (for $x<0$ and $x>L$) we set $V(\bm{r})=-U_\infty$
and take the limit $U_\infty\to\infty$.
For simplicity we set $\mu=0$ in the leads and we also assume that the magnetic
field is zero in the leads (so $\bm A$ is constant there). The Dirac equation
\begin{equation}
H\Psi=E\Psi\label{dirac}
\end{equation}
has to be solved subject to boundary conditions on the wave function $\Psi(\bm{r})$
at $y=0$ and $y=W$. We will 
consider two types of boundary conditions which mix neither valleys nor
transverse modes. The first is the periodic boundary condition 
$\left.\Psi\right|_{y=0}=\left.\Psi\right|_{y=W}$. The second is the infinite-mass 
boundary condition\footnote{Infinite mass boundary conditions are obtained by sending the mass
to infinity for $y<0$ and $y>W$. Particles are thus excluded from this region, much as an
infinite potential excludes Schr\"odinger particles. As a result the boundary condition of Eq.~(\ref{infmassbound})
is imposed at the boundaries $y=0$ and $y=W$  between the finite
(or zero) mass and the infinite mass regions. For more details, see Ref.~\onlinecite{Ber87}.} 
\begin{equation}
\label{infmassbound}
\left.\Psi\right|_{y=0}=\sigma_x\left.\Psi\right|_{y=0},\hspace{5mm}\left.\Psi\right|_{y=W}=-\sigma_x\left.\Psi\right|_{y=W}.
\end{equation}

We consider a scattering state $\Psi_n$ that has unit incident current from the left in mode $n$
and zero incident current from the right. (The quantum number $n$ labels transverse modes.)
In the leads $\Psi_n$ has the form
\begin{subequations}
\begin{align}
\Psi_n(\bm{r})&=\chi_n^+(y)\ e^{ik_n x}
+\sum_{m}r_{mn}\ \chi^{-}_m(y)\ e^{-ik_m x},\hspace{3mm}x<0,\\
\Psi_n(\bm{r})&=\sum_{m}t_{mn}\ \chi^{+}_m(y)\ e^{ik_m (x-L)},\hspace{5mm}x>L.
\end{align}
\end{subequations}
We have introduced
transmission and reflection amplitudes $t_{mn}$ and $r_{mn}$ and 
the longitudinal component $k_n$ of the wave vector of mode $n$. 
The right-propagating component in mode $n$ has a spinor $\chi_n^+$
and the left-propagating component has a spinor $\chi_n^-$. 

In the limit $U_\infty\to\infty$, the form of the scattering state in the leads can
be simplified considerably. The $n$-dependence of $k_n$ can be neglected, since
$k_n\simeq U_\infty/\hbar v\to \infty$ as $U_\infty \to \infty$. 
The number
$N_\infty\simeq U_\infty W/\hbar v$ of propagating modes in the leads can be
taken infinitely large. When $N_\infty\to\infty$, the choice of boundary condition in
the leads (not in the sample) becomes irrelevant and we choose periodic boundary conditions
in the leads for simplicity. 
Modes that are responsible for transport through the weakly doped sample have transverse
momenta $|q_n|\ll U_\infty$.
The corresponding spinors $\chi_n^\pm$ are
\begin{equation}
\chi^\pm_n(y)=\frac{1}{\sqrt{2W}}e^{iq_ny}\left(\begin{array}{rr}1\\\pm1\end{array}\right),\hspace{3mm}q_n=\frac{2\pi n}{W},\label{uinf}
\end{equation}
with $n=0,\,\pm1,\,\pm2,\,\ldots$.
While it is important not to neglect the finiteness of $q_n$ in the phase factor $\exp(iq_n y)$ of these modes,
the spinor structure is proportional to $(1,\pm1)$ independent of $n$, because $q_n/U_\infty\to0$.
We note the orthogonality relation
\begin{eqnarray}
\int_0^W dy\ \chi_m^\sigma(y)^\dagger\chi_n^{\sigma'}(y)&=&\delta_{m,n}\delta_{\sigma,\sigma'}.\label{comp1}
\end{eqnarray}
We also note that the definition of $\chi^\pm_n(y)$ ensures that each scattering state $\Psi_n$ carries
unit incident current.

In a similar way, we can define a scattering state incident from the right in mode $n$
with transmission and
reflection amplitudes $t'_{mn}$ and $r'_{mn}$. 
The transmission and reflection amplitudes constitute the scattering matrix
\begin{equation}
S=\left(\begin{array}{rr}r&t'\\t&r'\end{array}\right)\label{scateq},
\end{equation}
which is a unitary matrix that determines transport properties. For example, the conductance $G$
follows from the Landauer formula
\begin{equation}
G=\frac{4e^2}{h}{\rm Tr}\,tt^\dagger=\frac{4 e^2}{h}{\rm Tr}\, t'{t'}^\dagger,\label{landauer}
\end{equation}
where the factor of $4$ accounts for spin and valley degeneracies.

\subsection{Transfer matrix}
\label{transsect}
The information contained in the scattering matrix $S$ can equivalently be represented
by the transfer matrix $T$. While the scattering matrix relates outgoing waves to incoming waves,
the transfer matrix relates waves at the right,
\begin{equation}
\Psi_R(\bm{r})=\sum_{n,\sigma} b_n^\sigma\chi_n^\sigma(y)e^{i\sigma k_n(x-L)},~x>L,\label{psir}
\end{equation}
to waves at the left,
\begin{equation}
\Psi_L(\bm{r})=\sum_{n,\sigma} a_n^\sigma\chi_n^\sigma(y)e^{i\sigma k_nx},~x<0.\label{psil}
\end{equation}
The relation takes the form
\begin{equation}
b^\sigma_m=\sum_{n,\sigma'} T^{\sigma,\sigma'}_{m,n} a^{\sigma'}_n.\label{trans}
\end{equation}
The four blocks $T^{\sigma,\sigma'}$ of the transfer matrix are related to the 
transmission and reflection matrices by
\begin{subequations}
\label{ttos}
\begin{eqnarray}
r&=&-\left(T^{--}\right)^{-1} T^{-+},\label{ttosa}\\
t&=&T^{++}-T^{+-}\left(T^{--}\right)^{-1} T^{-+},\label{ttosb}\\
t'&=&\left(T^{--}\right)^{-1},\label{totsc}\\
r'&=&T^{+-}\left(T^{--}\right)^{-1}.\label{ttosd}
\end{eqnarray}
\end{subequations}
Unitarity of $S$ implies for $T$ the current conservation relation
\begin{equation}
T^{-1}=\Sigma_z T^\dagger \Sigma_z,
\label{current}
\end{equation}
where $\Sigma_z$ is a matrix in the space of modes with entries 
$\left(\Sigma_z\right)_{m,n}=\delta_{m,n}\,\sigma_z$ that are themselves $2\times 2$ matrices.
In terms of the transfer matrix the Landauer formula (\ref{landauer}) can be written as
\begin{equation}
G=\frac{4 e^2}{h}{\rm Tr}\left[\left({T^{--}}^\dagger T^{--}\right)^{-1}\right].\label{gdirac}
\end{equation}

\subsection{Real-space formulation}
In order to make contact with the network model, it is convenient to
change from the basis of transverse modes (labeled by the quantum number $n$)
to a real space basis (labeled by the transverse coordinate $y$). The real space transfer
matrix $X_{y,y'}$ is defined by
\begin{equation}
\Psi(L,y)=\int_0^W dy'\, X_{y,y'}\Psi(0,y')\label{eqX},
\end{equation}
where $\Psi(x,y)$ is any solution of the Dirac equation (\ref{dirac}) at
a given energy $E$. The kernel $X_{y,y'}$ is a $2\times2$ matrix, acting on the 
spinor $\Psi$. Because the integral (\ref{eqX}) extends only over the weakly doped
region, $X$ does not depend on the potential $U_\infty$ in  the leads. 

In view of the orthogonality relation (\ref{comp1}) the
real-space transfer matrix $X$ is related to the transfer matrix $T$
defined in the basis of modes in the leads by a projection onto $\chi_m^\pm$, 
\begin{equation}
T^{\sigma,\sigma'}_{m,n}=\int_0^Wdy\int_0^Wdy'\,\chi^\sigma_m(y)^\dagger X_{y,y'}\chi^{\sigma'}_n(y').\label{eqt1}
\end{equation}
We now substitute the explicit form of $\chi_n^\sigma$ from Eq.~(\ref{uinf}). 
The integrals over $y$ and $y'$ in Eq.~(\ref{eqt1}) amount to a Fourier transform,
\begin{equation}
X_{m,n}=\frac{1}{W}\int_0^Wdy\,\int_0^Wdy'\,e^{-iq_my}X_{y,y'}e^{iq_ny'}.
\end{equation}

From Eq.~(\ref{eqt1}) we conclude that the
$2\times2$ matrix structure of the transfer matrix, 
\begin{equation}
T_{m,n}=\left(\begin{array}{rr}T_{m,n}^{++}&T_{m,n}^{+-}\\T_{m,n}^{-+}&T_{m,n}^{--}\end{array}\right),
\end{equation}
is related to the $2\times2$ matrix structure of the real-space transfer matrix by a Hadamard transformation:
\begin{equation}
T_{m,n}={\cal H}X_{m,n}{\cal H},\hspace{3mm}
{\cal H}=\frac{1}{\sqrt{2}}\left(\begin{array}{rr}1&1\\1&-1\end{array}\right).
\label{xtot}
\end{equation}
(The unitary and Hermitian matrix $\cal H$ is called the Hadamard matrix.)
In view of Eq.~(\ref{current}), the current conservation relation for $X$ reads
\begin{equation}
X^{-1}=\Sigma_x X^\dagger\Sigma_x,\hspace{3mm}\left(\Sigma_x\right)_{m,n}=\delta_{m,n}\sigma_x,\label{current2}
\end{equation}
where we used ${\cal H}\sigma_z{\cal H}=\sigma_x$.
 
\section{Formulation of the network model}
\label{sect3}
\begin{figure}[t]
  \begin{center}
    \includegraphics[width=.7 \columnwidth]{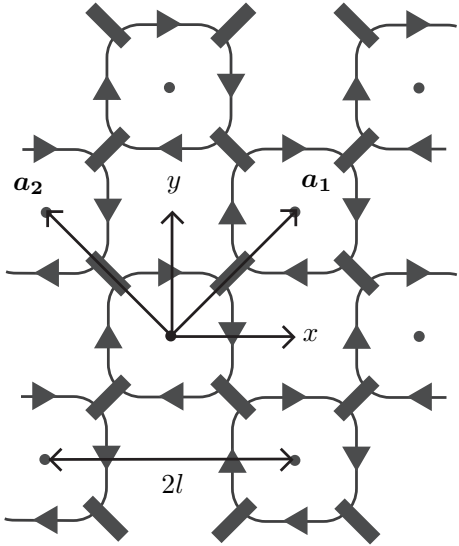}
    \caption{Square lattice (dots), with circulating current loops that form the
network model. The loops are coupled to nearest neighbors at the black rectangles.
The lattice vectors $\bm{a}_1$ and $\bm{a}_2$ (each of length $\sqrt{2}\,l$) are indicated. 
\label{networkfig}} 
  \end{center}
\end{figure}
The Chalker-Coddington network model \cite{Cha88,Kra05} was originally introduced in order
to analyze the localization transition in the quantum Hall effect. Our interest in this model
stems from the fact that it is known to map onto the two-dimensional Dirac equation.\cite{Ho96}
We briefly recall how the network model is defined and how the
mapping to the Dirac equation works.
We consider the square lattice shown in Fig.~\ref{networkfig},
with lattice constant $\sqrt{2}\,l$ and lattice vectors
\begin{equation}
\bm{a}_1=l(\bm{\hat x}+\bm{\hat y}),\hspace{5mm}\bm{a}_2=l(\bm{\hat y}-\bm{\hat x}). \label{basisvec}
\end{equation}
The integers $(m,n)$ label the lattice site $\bm{r}_{m,n}=m\bm{a}_1+n\bm{a}_2$. 
With each site is associated a single current loop
circling the site without enclosing any neighboring sites, 
say clockwise if viewed from the positive $\bm{z}$ axis. 
The radii of these loops are expanded until states associated with
nearest neighboring sites overlap. 
At these points of overlap, states on adjacent loops can scatter into
each other. 

\begin{figure}[ht]
  \begin{center}
    \includegraphics[width=.70 \columnwidth]{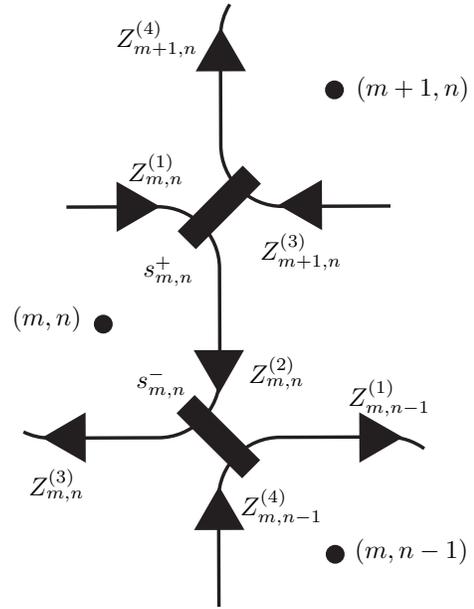}
    \caption{Segment of the network of Fig.\ \ref{networkfig} with
the wave amplitudes $Z_{m,n}^{(n)}$ and scattering matrices $s_{m,n}^\pm$ indicated.
\label{scatfig}} 
  \end{center}
\end{figure}

As illustrated in Fig.~\ref{scatfig}, four current amplitudes
$Z^{(k)}_{m,n}$, $k=1,\ldots,4$ are associated with each site $(m,n)$.
These are amplitudes incident upon points of overlap, ordered clockwise,
starting from the point of overlap with site $(m+1,n)$. Each incident wave amplitude
$Z_{m,n}^{(k)}$ has picked up a phase $\phi_{m,n}^{(n)}$ since the previous point of overlap.
With the point of overlap between loop $(m,n)$ and $(m+1,n)$ is associated a $2\times 2$ scattering matrix 
$s^{+}_{m,n}$, while $s^{-}_{m,n}$ is associated with the point of overlap between $(m,n)$ and $(m,n-1)$.

The matrix elements of $s^{+}_{m,n}$ and $s^{-}_{m,n}$ are arranged such that
\begin{subequations}
\label{networkdef}
\begin{eqnarray}
\left(\begin{array}{c}
Z^{(2)}_{m,n}\\Z^{(4)}_{m+1,n}\end{array}\right)&=&\left(\begin{array}{cc}e^{i\phi^{(2)}_{m,n}}&0\\
0&e^{i\phi^{(4)}_{m+1,n}}\end{array}\right)s^{+}_{m,n}
\left(\begin{array}{c}Z^{(1)}_{m,n}\\Z^{(3)}_{m+1,n}\end{array}\right),\nonumber\\
&&\label{neteq1}\\
\left(\begin{array}{c}
Z^{(1)}_{m,n-1}\\Z^{(3)}_{m,n}\end{array}\right)&=&\left(\begin{array}{cc}e^{i\phi^{(1)}_{m,n-1}}&0\\
0&e^{i\phi^{(3)}_{m,n}}\end{array}\right)s^{-}_{m,n}
\left(\begin{array}{c}Z^{(2)}_{m,n}\\Z^{(4)}_{m,n-1}\end{array}\right).\nonumber\\
&&\label{neteq2}
\end{eqnarray}
\end{subequations} 

Ho and Chalker\cite{Ho96} showed how this model can be mapped onto 
the Dirac equation for two-dimensional fermions.
Firstly, one parametrizes the scattering matrices $s_{m,n}^{\pm}$ in terms of Pauli matrices $\sigma_i$,
\begin{subequations}
\begin{eqnarray}
s^{-}_{m,n}&=&\sin\left(\frac{\pi}{4}+\beta_{m,n}\right)\sigma_z+\cos\left(\frac{\pi}{4}+\beta_{m,n}\right)\sigma_x,\nonumber\\
&&\\
s^{+}_{m,n}&=&\cos\left(\frac{\pi}{4}+\beta_{m,n}\right)\sigma_z+\sin\left(\frac{\pi}{4}+\beta_{m,n}\right)\sigma_x.\nonumber\\
&&
\end{eqnarray}
\end{subequations}
(The same matrix of coefficients $\beta_{m,n}$ is used for $s^+_{m,n}$ and $s^-_{m,n}$.) For given fields $V(\bm{r})$, 
$\bm{A}(\bm{r})$, and $\mu(\bm{r})$ in the Dirac equation, the mapping then dictates a corresponding choice of
parameters in the network model, namely $\phi_{m,n}^{(k)}$ and $\beta_{m,n}$ have to satisfy\cite{Ho96}
\begin{subequations}
\label{corresp}
\begin{eqnarray}
\frac{1}{2}\sum_{k=1}^4\phi^{(k)}_{m,n}&=&\left[E-V(\bm{r}_{m,n})\right]\frac{l}{\hbar v},\\
\frac{\phi_{m,n}^{(1)}-\phi^{(3)}_{m,n}}{2}&=&A_x(\bm{r}_{m,n})\frac{e l}{\hbar v},\\
\frac{\phi_{m,n}^{(4)}-\phi^{(2)}_{m,n}}{2}&=&A_y(\bm{r}_{m,n})\frac{e l}{\hbar v},\\
2\beta_{m,n}&=&\mu(\bm{r}_{m,n})\frac{l}{\hbar v}.
\end{eqnarray}
\end{subequations}
With this choice of parameters there is an approximate equality between a solution $\Psi(\bm{r})$ of
the Dirac equation and the current amplitudes of the network model,
\begin{equation}
\Psi(\bm{r}_{m,n})\approx {\cal G}\left(\begin{array}{c}Z^{(1)}_{m,n}\\Z^{(3)}_{m,n}\end{array}\right),\hspace{3mm}
{\cal G}=\frac{1}{\sqrt{2}}\left(\begin{array}{rr}1&i\\1&-i\end{array}\right).\label{connection}
\end{equation}
The accuracy of the approximation is improved by making the lattice constant $\sqrt{2}\,l$ smaller and smaller.

As mentioned in Sec.~\ref{sect2}, we will be considering two types of boundary conditions at 
$y=0$ and $y=W$ in the sample region $0<x<L$.
The periodic boundary condition is realized in the network model by putting the square
lattice on a cylinder of circumference $W=2Nl$ oriented along the $x$-axis. The infinite-mass
boundary condition is realized\cite{Ho96} by terminating the square lattice at $y=0$ and $y=W$ and adjusting the 
scattering phases along the edge. 
The edge $y=0$ lies at sites $(n,-n)$ and the edge $y=W$ lies at sites
$(N-1+n,N-1-n)$. 
As shown in App.~\ref{bounds}, for sites $(n,-n)$ Eq.~(\ref{networkdef}) must be replaced with 
\begin{equation}
Z^{(4)}_{n,-n}=-Z^{(3)}_{n,-n},\hspace{3mm}Z^{(3)}_{n,-n}=Z^{(2)}_{n,-n},
\end{equation}
while for sites $(N+n,N-n)$ it must be replaced with
\begin{equation}
Z^{(2)}_{N+n,N-n}=Z^{(1)}_{N+n,N-n},\hspace{3mm}Z_{N+n,N-n}^{(4)}=Z_{N+n,N-n}^{(1)}.
\end{equation}
 
\section{Correspondence between scattering matrices of Dirac equation and network model}
\label{sect4}
In this section we combine the  known results summarized in the previous two sections to
construct the scattering matrix $S$ of a graphene strip with heavily doped leads from a 
solution of the network model. This construction does not immediately follow from the correspondence
(\ref{connection}) because the limit $U_\infty\to\infty$ of heavily doped leads still needs 
to be taken. At first glance it would seem that, in order to preserve the correspondence
between the network model and the Dirac equation, we must simultaneously take the
limit $l\to0$ so that $U_\infty l/\hbar v$ remains small. (The correspondence between
the network model and the Dirac equation is correct only to first order in this quantity.) 
This would imply that very large networks are required for an accurate representation
of the graphene strip.

It turns out, however, that it is not necessary to model the heavily doped leads explicitly 
in the network model, as we now demonstrate. We define the real-space transfer matrix $Y$ 
as the matrix that relates $Z^{(1)}$ and $Z^{(3)}$ at the right edge of the network to 
$Z^{(1)}$ and $Z^{(3)}$ at the left edge of the network. The left edge ($x=0$) lies at sites
$(n,n)$ with $n=0,\,1,\,2,\,\ldots,N-1$. The right edge at $x=L=2Ml$ lies at sites $(n+M,n-M)$.
The real-space transfer matrix $Y$ relates
\begin{equation}
\label{defy}
\left(\begin{array}{r}Z^{(1)}_{n+M,n-M}\\Z^{(3)}_{n+M,n-M}\end{array}\right)
=\sum_{n'=0}^{N-1}Y_{n,n'}
\left(\begin{array}{r}Z^{(1)}_{n',n'}\\Z^{(3)}_{n',n'}\end{array}\right).
\end{equation}
We define the Fourier transform
\begin{equation}
Y_{q_m,q_n}=\frac{1}{N}\sum_{m'=0}^{N-1}\sum_{n'=0}^{N-1} e^{-2 i l q_m m'}Y_{m',n'}e^{2 i l q_n n'},
\end{equation}
with $q_n=2\pi n/W$.

In view of the relation (\ref{connection}) between the Dirac wave function $\Psi$ and the
network amplitudes $Z^{(1)}$, $Z^{(3)}$, the real space transfer matrix $X$ of the Dirac equation
is related to $Y$ by a unitary transformation,
\begin{equation}
X_{y=2ln,y'=2ln'}=\frac{1}{2l}\,{\cal G}\, Y_{n,n'}\, {\cal G}^\dagger.
\end{equation}
We can now use the relation (\ref{xtot}) between $X$ and the transfer matrix $T$ to obtain
\begin{equation}
T_{m,n}=\left(\begin{array}{rr}1&0\\0&i\end{array}\right)Y_{q_m,q_n}
\left(\begin{array}{rr}1&0\\0&-i\end{array}\right),\label{ttoy}
\end{equation}
where we have used
\begin{equation}
{\cal H}{\cal G}=\left(\begin{array}{rr}1&0\\0&i\end{array}\right).
\end{equation}

From Eq.~(\ref{ttoy}) it follows that the lower right blocks of $T$ and $Y$ are equal: 
$T^{--}_{m,n}=Y^{--}_{q_m,q_n}$. 
Substitution into the Landauer formula (\ref{gdirac}) gives
\begin{equation}
G=\frac{4e^2}{h}{\rm Tr}\left[\left({Y^{--}}^{\dagger}Y^{--}\right)^{-1}\right].\label{gnet}
\end{equation}
The Landauer formula applied to the network model thus gives the conductance of the 
corresponding graphene sheet connected to heavily
doped leads. For later use, we note the current conservation relation for $Y$, which
follows from Eqs.~(\ref{current}) and (\ref{ttoy})
\begin{equation}
Y^{-1}=\Sigma_zY^\dagger\Sigma_z.
\end{equation}

\section{Numerical Solution}
\label{five}
\begin{figure}[t]
  \begin{center}
    \includegraphics[width=.9 \columnwidth]{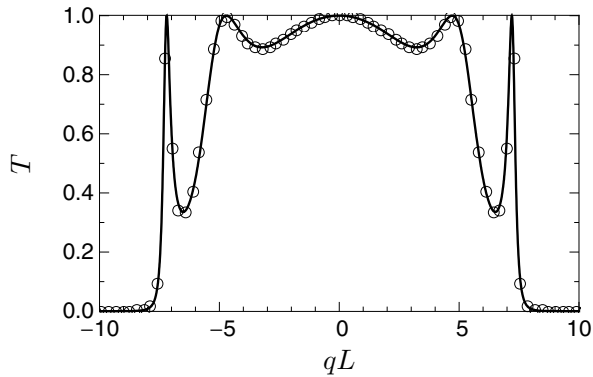}
    \caption{Transmission probability of a clean graphene sheet, at energy $E=7.85\,\hbar v/L$ as a
function of transverse wave number $q$. The solid line is the result (\ref{cleantrans}) from the Dirac 
equation, while the open circles were numerically calculated using the network model with periodic 
boundary conditions (when $q=2\pi n/W$). The discretization parameter of the network was 
$\epsilon=El/\hbar v=0.28$.\label{fig1}}
  \end{center}
\end{figure}
In this section we test the accuracy and efficiency of the solution of a
scattering problem in graphene by means of the network model. As explained in 
Sec.~\ref{sect4} we need to calculate the real space transfer matrix $Y$ through
the weakly doped region. The conductance of the corresponding graphene sample then follows
from Eq.~(\ref{gnet}).

We calculate the real-space transfer matrix recursively by adding slices to the network
and multiplying the transfer matrices of individual slices. 
Since a multiplication of transfer matrices is numerically unstable we
stabilize the algorithm as explained in App.~\ref{numapp}. We limit the numerical investigation
in this section to the case $A(\bm{r})=0$, $\mu(\bm{r})=0$ where only the electrostatic
potential $V(\bm{r})$ is non-zero.

We have found that the efficiency of the algorithm can be improved by using the fact that,
according to Eq.~(\ref{corresp}), there is some arbitrariness in the choice of the phases 
$\phi^{(1)},\ldots,\phi^{(4)}$. For $\bm{A}(\bm{r})=0$ and
$\mu(\bm{r})=0$, one choice of the phases could be
\begin{equation}
\phi^{(k)}_{m,n}=\left[E-V(m\bm{a}_1+n\bm{a}_2)\right]l/2, \hspace{3mm}k=1,\ldots,4.
\end{equation}
Another choice is
\begin{equation}
\label{choice2}
\phi^{(1)}_{m,n}=\phi^{(3)}_{m,n}=\left[E-V(\bm{r}_{m,n})\right]l,\hspace{3mm}
\phi^{(2)}=\phi^{(4)}=0.
\end{equation}
The correspondence (\ref{connection}) between the network model and the Dirac equation holds
for both choices of the phases, however the corrections for finite $l$ are 
smaller for choice (\ref{choice2}). More precisely, as shown in App.~\ref{phase}, if $\phi^{(2)}$
and $\phi^{(4)}$ are zero, the network model does not contain corrections to the Dirac equation of 
order $\bm{\partial_r}Vl$. 

Let us first consider 
the analytically solvable case of a clean graphene sheet that is obtained by
setting $V=0$ in the weakly doped region.
The Dirac equation gives transmission probabilities\cite{Two06}
\begin{subequations}
\label{cleantrans}
\begin{eqnarray}
T(E,q)&=&\left|\cos\xi L+i\frac{E\sin{\xi L}}{\hbar v\xi}\right|^{-2},\\
\xi&=&\sqrt{\left(\frac{E}{\hbar v}\right)^2-q^2}.
\end{eqnarray}
\end{subequations}
For periodic boundary conditions the transverse wave vector is discretized as
$q_n=2\pi n/W,$ with $n=0,\,\pm1,\,\pm2,\,\ldots$

In Fig.~\ref{fig1} we compare Eq.~(\ref{cleantrans}) to the results from the network model for 
periodic boundary conditions in the weakly doped region. The small parameter that controls the accuracy
of the correspondence is $\epsilon=El/\hbar v$. We find excellent agreement for a relatively large 
$\epsilon\simeq0.3$.

\begin{figure}[t]
  \begin{center}
    \includegraphics[width=.9 \columnwidth]{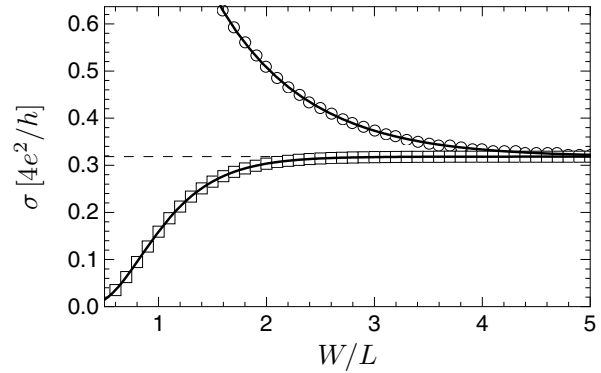}
    \caption{Conductivity $\sigma=G\times L/W$ at $E=0$ for a clean graphene sheet as a function
of the aspect ratio. The data points were calculated from the network model for fixed $L=40\,l$
with periodic boundary conditions (circles) and infinite mass boundary conditions (squares)
in the weakly doped region. The solid lines are the result\cite{Two06} from the Dirac equation.
The dashed line indicates the limiting value $\sigma h/4 e^2=1/\pi$ for short wide samples.  
\label{boundconfig}}
  \end{center}
\end{figure}

Fig.~\ref{boundconfig} shows the conductivity
\begin{equation}
\sigma=\frac{L}{W}\frac{4e^2}{h}\sum_n T(E,q_n)
\end{equation}
at the Dirac point ($E=0$) as a 
function of the aspect ratio $W/L$. We do the calculation both for periodic and infinite mass boundary conditions
in the weakly doped region. (In  the latter case $q_n=(n+\tfrac{1}{2})\pi/W$ with $n=0,\,1,\,2,\ldots$)
Again we see excellent agreement with the analytical results from the Dirac equation\cite{Two06}. 

\begin{figure}[t]
  \begin{center}
    \includegraphics[width=.9 \columnwidth]{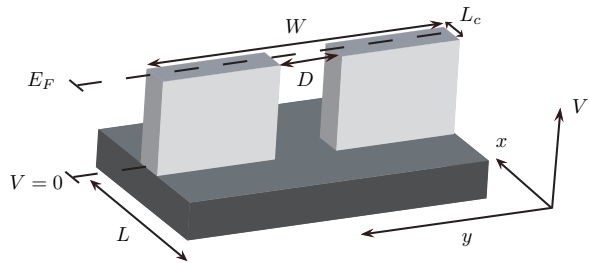}
    \caption{Potential landscape $V(x,y)$ that produces a quantum point contact.
 The Fermi energy $E_F$ is indicated.\label{setupfig}}
  \end{center}
\end{figure}
We now apply the network model to a case that cannot be solved analytically, because it involves inter-mode 
scattering. We take the electrostatic potential landscape shown in Fig.~\ref{setupfig}, which produces a narrow
constriction or quantum point contact of width $D$ and length $L_c$.
In the weakly doped region, of length $L$, electrons have an energy $E_F$ measured from the Dirac point. The barrier
potential is tuned so that electron transport through the barrier takes place at the Dirac point, where all
waves are evanescent.  
As the constriction is widened, the number of modes at a given energy that propagates through the opening increases.
For fixed $E_F$, this should lead to steps in the conductance as a function of opening width, at 
intervals of roughly $\pi/E_F$.
The steps are smooth because the current can also tunnel through the barrier.

\begin{figure}[t]
  \begin{center}
    \includegraphics[width=.9 \columnwidth]{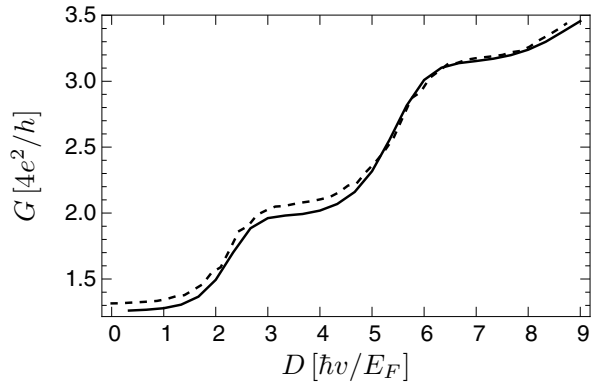}
    \caption{Conductance through the constriction of Fig.~\ref{setupfig} as a function of the width
of the opening in the constriction. The solid line was obtained using the network model,
while the dashed line was obtained using the tight-binding model of graphene. 
We used parameters $W=35\, \hbar v/E_F$, $L_c=8.7\, \hbar v/E_F$. 
For the network model we set the length of the weakly doped region to $L=49\, \hbar v/E_F$ and used 
a lattice constant 
$\sqrt{2} l=0.24\, \hbar v/E_F$, while in the tight-binding calculation we used
a lattice constant $0.17\, \hbar v/E_F$.\label{netqpc} } 
  \end{center}
\end{figure}

We have calculated the conductance with the network model (solid curve in Fig.~\ref{netqpc}) and using
the tight-binding model of graphene (dashed curve). In the tight-binding calculation we did not connect
heavily doped leads to the weakly doped region. 
This does not affect the results, as long as $L\gg L_c$. 

Both calculations show a smooth 
sequence of steps in the conductance. The agreement is reasonably good, but not as good as in the previous cases.
This can be understood since the tight-binding model of graphene is only
equivalent to the Dirac equation on long length-scales.

The final numerical study that we report on in this paper involves transport
at the Dirac point through a disordered potential landscape. Recent experimental
studies\cite{Mar08} have observed electron and hole puddles in undoped graphene.
The correlation length of the potential is larger than the lattice constant, hence intervalley scattering is weak. We are therefore in the regime of applicability of the network model (which eliminates intervalley scattering from the outset).

\begin{figure}[ht]
\begin{center}
\includegraphics[width=.95 \columnwidth]{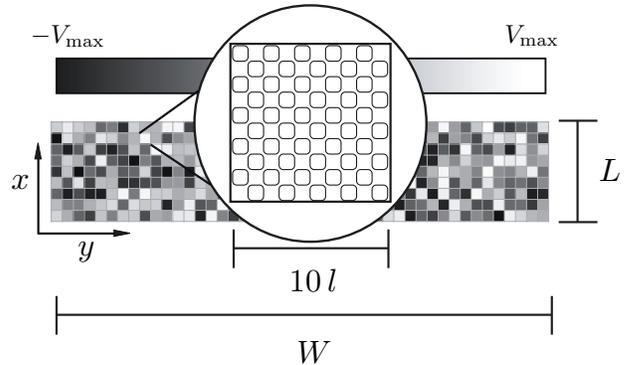}
\caption{Illustration of the model of electron and hole puddles in a graphene strip that we have studied. The sample is divided into
tiles. The value of the potential on a tile is a constant, here indicated in gray-scale,
uniformly distributed between $-V_{\rm max}$ and $V_{\rm max}$. The potential on
different tiles is uncorrelated. We choose a mesh for the network such that each
tile has size $10\,l\times 10\,l$, where the network lattice constant is $\sqrt{2} l$.\label{figpot}}
\end{center}
\end{figure}
To model the electron and hole puddles, we devide the sample into an 
array of square tiles (Fig~\ref{figpot}),
where each tile has size $10\,l\times10\,l$, $\sqrt{2}l$ being the lattice constant of
the network model. The electrostatic potential
is constant on a single tile, but uncorrelated with the potential on the other
tiles. We take the values of the potential on any given tile to be a random variable
uniformly distributed between $-V_{\rm max}$ and $V_{\rm max}$. To make contact
with previous studies\cite{Ryc07,Bar07}, we quantify the disorder strength by the dimensionless number
\begin{equation}
K_0=\frac{1}{(\hbar v)^2}\int d\bm{r'}\left<V(\bm{r})V(\bm{r'})\right>.
\end{equation}
(The average $\left<V(\bm{r})\right>$ is zero.)
With tiles of dimension $10\,l\times10\,l$, 
the relation between $K_0$ and $V_{\rm max}$ is $K_0=100 (V_{\rm max} l/\hbar v)^2/3$
and the network model faithfully represents the Dirac equation
for values up to $K_0\simeq10$.
We use a sample with aspect ratio $W/L=5$ and average over $100$ disorder realizations.
We repeat the calculation for two different sample sizes namely $W=5L=300l$ and $W=5L=450l$. The calculation
is performed for transport at energy $E=0$, i.e. the Dirac point of a clean, undoped sample.
\vspace*{2mm}
\begin{figure}[ht]
\begin{center}
\includegraphics[width=.98 \columnwidth]{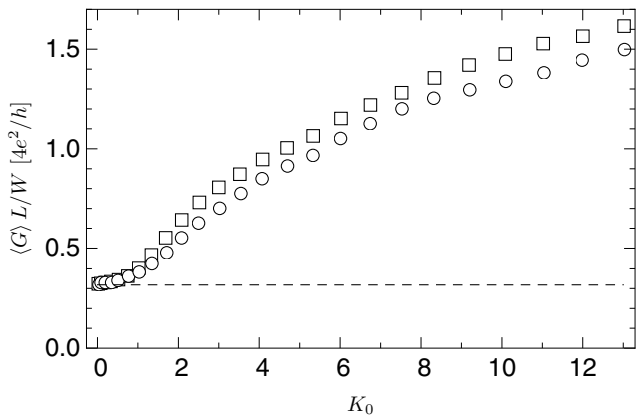}
\caption{Conductivity $\sigma=G\,L/W$ averaged over $100$ disorder realizations versus disorder strength $K_0$
at the Dirac point $E=0$.
The circles are for samples of size $60\,l\times300\,l$ while squares are for samples of size $90\,l\times450\,l$.
The statistical error is of the order of the size of the data points.
The dotted line indicates the ballistic limit $G\,L/W=4e^2/\pi h$.\label{avecond}}
\end{center}
\end{figure}
In Fig.~\ref{avecond} we show the average conductance. 
Remakably enough the conductance increases with increasing disorder strength. This is consistent
with the results obtained in Refs. \onlinecite{Ryc07,Bar07,Nom07,San07}. The effect should not depend
on the shape of the tiles in our model for the disorder. 
We have therefore repeated the calculation with rhombic instead of square tiles.
We find deviations of less than $5\%$.

The increase in conductance is explained by
the non-zero density of states at the Dirac point that is induced by the disorder, together with the
absence of back-scattering for Dirac electrons. While we do not make a detailed study of the dependence
of conductance on sample size (at fixed aspect ratio), we note that the conductance of larger samples
(squares in Fig.~\ref{avecond}) is larger than the conductance of the smaller samples 
(circles in Fig.~\ref{avecond}). This is consistent with the scaling behavior found in Refs.~\onlinecite{Bar07,Nom07,San07}.

\section{Conclusion}
\label{conclusion}
In conclusion, we have shown how the Chalker-Coddington network model can be 
used to solve a scattering problem in a weakly doped graphene sheet between
heavily doped electron reservoirs (which model the metallic contacts). 
The method is particularly useful when the scattering problem does not allow an analytical
solution, so that a numerical solution is required. The network model
eliminates intervalley scattering from the outset. Thus, with a given mesh size,
a larger graphene sample can be modeled with the network model than with methods
based on the honeycomb lattice. The key technical result of our work is that 
an infinitely high potential step at the contacts can be implemented
analytically by a unitary transformation of the real-space transfer matrix, without 
having to adjust the lattice constant of the network model to the
small values needed to accommodate the small wave length in the contacts. 
We have demonstrated that the algorithm provides an accuracy and efficiency
comparable to the tight-binding model on a honeycomb lattice. In agreement with
the existing literature\cite{Ryc07,Bar07,Nom07,San07} we have found that disorder that is smooth on the scale of the
graphene lattice constant enhances conductivity at the Dirac point. The absence of 
intervalley scattering in the network model may prove useful for the study
of these and other single-valley properties.

\acknowledgments
This research was supported by the Dutch Science Foundation NWO/FOM and
by the European Union's Marie Curie Research Training Network (contract
MRTN-CT-2003-504574, Fundamentals of Nano-electronics).

\appendix

\section{Infinite-mass boundary condition for the network model}
\label{bounds}
\begin{figure}[t]
  \begin{center}
    \includegraphics[width=.75 \columnwidth]{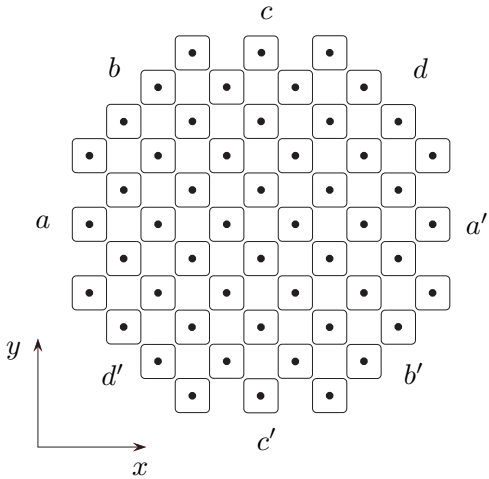}
    \caption{Network of circulating current loops, as in Fig.~\ref{networkfig}, but now terminated with straight
edges. The letters $a$, $b,\,\ldots$ label the orientation of the edge.\label{boundaries}}
  \end{center}
\end{figure}

In this appendix we consider the boundary condition imposed on the Dirac equation
by termination of the network along a straight edge. We consider the eight orientations
shown in Fig.~\ref{boundaries} which have the shortest periodicity along the edge.
Since we want to discuss the long wave-length limit,
each edge needs to be much longer than the lattice constant $\sqrt{2}l$. (In this respect the figure with
its relatively short edges is only schematic.) 
The orientations are defined by the vector
$\bm{\hat{n}}(\alpha)=-\bm{\hat{x}}\sin\alpha+\bm{\hat{y}}\cos\alpha$,
 $\alpha=j\pi/4$, $j=1,\ldots,8$ which is perpendicular to the edge and points outwards. 

We wish to impose the infinite mass boundary condition\cite{Ber87}
\begin{eqnarray}
\Psi_{\rm edge}&=&[\bm{\hat{n}}(\alpha)\times\bm{\hat{z}}]\cdot\bm{\sigma}\,\Psi_{\rm edge}\nonumber\\
&=&(\sigma_x\cos\alpha+\sigma_y\sin\alpha)\Psi_{\rm edge}
\label{infmass}
\end{eqnarray}
on the Dirac wavefunction at the edge. 
In view of the correspondence (\ref{connection}) between the Dirac equation and the network model, 
Eq.~(\ref{infmass}) implies the boundary condition
\begin{equation}
\label{infmass2}
\left(\begin{array}{c}Z^{(1)}\\Z^{(3)}\end{array}\right)_{\rm edge}
=\left(-\sigma_x\,\sin\alpha+\sigma_z\,\cos\alpha\right)\left(\begin{array}{c}Z^{(1)}\\Z^{(3)}\end{array}\right)_{\rm edge}
\end{equation}
on the network amplitudes.


Away from the edge, the network amplitudes obey the equations (\ref{networkdef}). For $\mu$, $\bm{A}$,
$V$, and $E$ all equal to zero (Dirac point) these reduce to
\begin{subequations}
\label{dpoint}
\begin{eqnarray}
\left(\begin{array}{c}
Z^{(2)}_{m,n}\\Z^{(4)}_{m+1,n}\end{array}\right)&=&
{\cal H}\left(\begin{array}{c}Z^{(1)}_{m,n}\\Z^{(3)}_{m+1,n}\end{array}\right)\label{dpoint1},\\
\left(\begin{array}{c}
Z^{(1)}_{m,n-1}\\Z^{(3)}_{m,n}\end{array}\right)&=&
{\cal H}\left(\begin{array}{c}Z^{(2)}_{m,n}\\Z^{(4)}_{m,n-1}\end{array}\right)\label{dpoint2}.
\end{eqnarray}
\end{subequations} 
We can eliminate the amplitudes $Z^{(2)}$ and $Z^{(4)}$ to arrive at the equations
\begin{subequations}
\label{bulk}
\begin{align}
Z^{(1)}_{m,n}={}&\frac{1}{2}\big[Z^{(1)}_{m,n+1}+Z^{(1)}_{m-1,n}\nonumber\\
&\hspace{1.7cm}-Z_{m,n}^{(3)}+Z_{m+1,n+1}^{(3)}\big]\label{bulk1}\\
Z^{(3)}_{m,n}={}&\frac{1}{2}\big[Z^{(1)}_{m,n}-Z^{(1)}_{m-1,n-1}\nonumber\\
&\hspace{1.7cm}+Z_{m+1,n}^{(3)}+Z_{m,n-1}^{(3)}\big].\label{bulk3}
\end{align}
\end{subequations}
There are two linearly independent solutions $(Z_{m,n}^{(1)},Z_{m,n}^{(3)})\propto(1,0)$
and $(Z_{m,n}^{(1)},Z_{m,n}^{(3)})\propto(0,1)$. When the network is truncated along an edge,
the bulk equations (\ref{bulk}) do not hold for the amplitudes along the edge. We seek the
modified equations that impose the boundary condition (\ref{infmass2}) up to corrections of order
$(E-V)l/\hbar v$.

The edge orientation $a$ was previously considered by Ho and Chalker\cite{Ho96}. We consider
here all four independent orientations $a$, $b$, $c$, and $d$. The other four orientations
$a'$, $b'$, $c'$, and $d'$ are obtained by a symmetry relation.

\begin{figure}[h]
  \begin{center}
    \includegraphics[width=.6 \columnwidth]{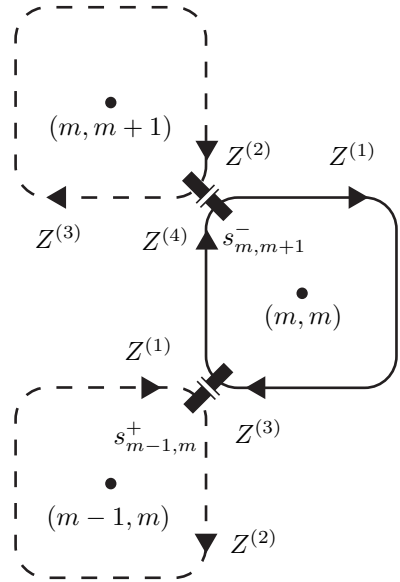}
    \caption{Network amplitudes at an edge with orientation $a$. The dashed current loops are 
     removed.\label{edgea}}
  \end{center}
\end{figure}

Edge $a$ is constructed by removing all sites $(m,n)$ with $n>m$. (See Fig.~\ref{edgea}.)
This means that
the network amplitudes $Z^{(3)}_{m,m}$ are prevented from scattering into the non-existent amplitudes
$Z^{(2)}_{m-1,m}$ belonging to the removed sites $(m-1,m)$. Similarly, the amplitudes $Z^{(4)}_{m,m}$
are prevented from scattering into the non-existent amplitudes $Z^{(3)}_{m,m+1}$. 
To do this one must modify the scattering matrices
$s_{m-1,m}^{+}$ so that $Z^{(3)}_{m,m}$ can only scatter into $Z^{(4)}_{m,m}$ and
$s_{m,m+1}^{-}$ so that $Z^{(4)}_{m,m}$ can only scatter into $Z^{(1)}_{m,m}$.
As a consequence, for $n=m+1$ Eq.~(\ref{dpoint}) is replaced by
\begin{equation}
Z^{(4)}_{m,m}=-Z^{(3)}_{m,m},\hspace{3mm}Z^{(1)}_{m,m}=Z^{(4)}_{m,m}.
\end{equation}
We eliminate $Z^{(2)}$ and $Z^{(4)}$ to arrive at Eq.~(\ref{bulk}) for $n<m$ and
Eq.~(\ref{bulk3}) for $n=m$. Eq.~(\ref{bulk1}) for $n=m$ is replaced by
\begin{equation}
Z_{m,m}^{(1)}=-Z^{(3)}_{m,m}.
\end{equation}
The solution 
$(Z^{(1)}_{m,n},Z^{(3)}_{m,n})\propto(1,-1)$ 
indeed satisfies the infinite mass boundary condition (\ref{infmass2}) with
$\alpha=\pi/2$.

\begin{figure}[h]
  \begin{center}
    \includegraphics[width=.6 \columnwidth]{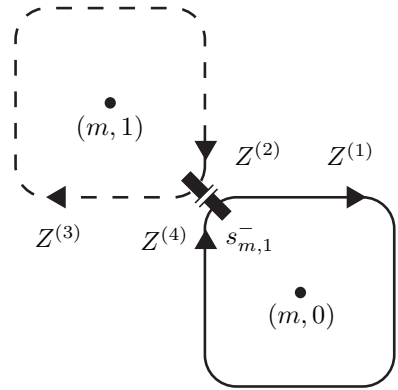}
    \caption{Edge with orientation $b$.\label{edgeb}}
  \end{center}
\end{figure}

Edge $b$ is constructed by removing all sites $(m,n)$ with $n>0$. (See Fig.~\ref{edgeb}.) 
This means that
the network amplitudes $Z^{(4)}_{m,0}$ are prevented from scattering into the non-existent amplitudes
$Z^{(3)}_{m,1}$ belonging to the removed sites $(m,1)$. 
For $n=1$, we replace Eq.~(\ref{dpoint2}) by 
\begin{equation}
Z^{(1)}_{m,0}=Z^{(4)}_{m,0}.
\end{equation}
If we now eliminate the amplitudes $Z^{(2)}$ and $Z^{(4)}$ we find that
Eq.~(\ref{bulk}) is still valid for all $n<0$. 
For $n=0$, Eq.~(\ref{bulk3}) still holds, while Eq.~(\ref{bulk1}) is changed to 
\begin{equation}
Z_{m,0}^{(1)}=\frac{1}{\sqrt{2}}\left(Z^{(1)}_{m-1,0}-Z^{(3)}_{m,0}\right).
\end{equation}
The solution
$(Z^{(1)}_{m,n},Z^{(3)}_{m,n})^{\rm T}\propto(1,1-\sqrt{2})$
satisfies the infinite mass boundary condition (\ref{infmass2}) with
$\alpha=\pi/4$.

\begin{figure}[h]
  \begin{center}
    \includegraphics[width=.9 \columnwidth]{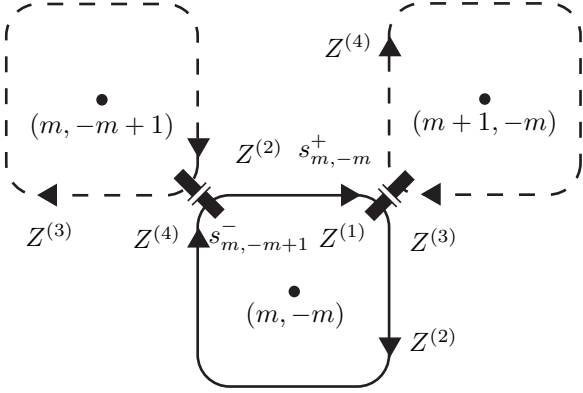}
    \caption{Edge with orientation $c$.\label{edgec}}
  \end{center}
\end{figure}

Next, we consider edge $c$, which results from the removal of all sites $(m,n)$ with $m>-n$. (See Fig.~\ref{edgec}.)
In this case, $s_{m,-m+1}^{-}$ must be modified to prevent $Z_{m,-m}^{(4)}$ from scattering into $Z_{m,-m+1}^{(3)}$.
Furthermore, $s_{m,-m}^{+}$ must be modified to prevent $Z_{m,-m}^{(1)}$ from scattering into $Z_{m+1,-m}^{(4)}$. 
For $n=-m+1$ we replace Eq.~(\ref{dpoint}) by
\begin{equation}
\label{a9}
Z^{(2)}_{m,-m}=Z^{(1)}_{m,-m},\hspace{3mm}Z_{m,-m}^{(1)}=Z_{m,-m}^{(4)},
\end{equation}
and eliminate $Z^{(2)}$ and $Z^{(4)}$ to verify that the boundary condition holds.

The condition (\ref{a9}) modifies three of the equations (\ref{bulk}):
\begin{subequations}
\begin{eqnarray}
Z^{(1)}_{m,-m}&=&\frac{1}{\sqrt{2}}\left(Z^{(1)}_{m-1,-m}-Z^{(3)}_{m,-m}\right),\\
Z^{(3)}_{m,-m}&=&\frac{1}{2}\big(Z^{(3)}_{m,-m-1}-Z^{(1)}_{m-1,-m-1}\nonumber\\
&&\hspace{1cm}+\hspace{.5mm}\sqrt{2}Z^{(1)}_{m,-m}\big),\\
Z^{(1)}_{m,-m-1}&=&\frac{1}{2}\big(-Z^{(3)}_{m,-m-1}+Z^{(1)}_{m-1,-m-1}\nonumber\\
&&\hspace{1cm}+\hspace{.5mm}\sqrt{2}Z^{(1)}_{m,-m}\big).\label{edge}
\end{eqnarray}
\end{subequations}
For $m<-n-1$ Eq.~(\ref{bulk}) holds without modification and Eq.~(\ref{bulk3}) also holds for $m=-n-1$. 
The solution 
\begin{equation}
Z^{(1)}_{m,n<-m}=\sqrt{2}Z^{(1)}_{m,-m}={\rm constant},\hspace{3mm}Z^{(3)}_{m,n}=0
\end{equation}
implies $(Z^{(1)}_{m,n},Z^{(3)}_{m,n})\propto(1,0)$ for $m<-n$, which satisfies the infinite mass 
boundary condition (\ref{infmass2}) with $\alpha=0$.

\begin{figure}[h]
  \begin{center}
    \includegraphics[width=.6 \columnwidth]{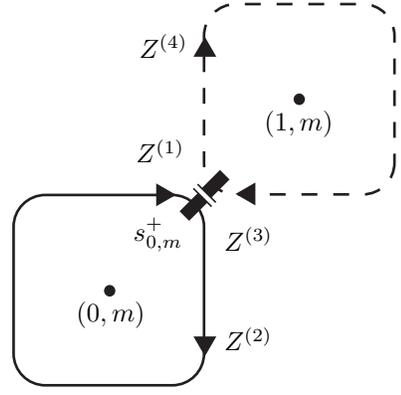}
    \caption{Edge with orientation $d$ . \label{edged}}
  \end{center}
\end{figure}

Edge $d$ results from the removal of all sites $(n,m)$ with $m>0$. (See Fig.~\ref{edged}.)
We must modify $s^{+}_{0,m}$ such that $Z^{(1)}_{0,m}$ does not scatter into 
$Z^{(4)}_{1,m}$. To do this we replace Eq.~(\ref{dpoint1}) for sites $(0,m)$ by 
\begin{equation}
Z^{(2)}_{0,m}=Z^{(1)}_{0,m}.
\end{equation}
We again eliminate $Z^{(2)}$ and $Z^{(4)}$ to arrive at
\begin{subequations}
\begin{eqnarray}
Z^{(1)}_{0,m}&=&\frac{1}{\sqrt{2}}\left(\sqrt{2}Z^{(1)}_{0,m+1}+Z_{-1,m}^{(1)}-Z_{0,m}^{(3)}\right),\nonumber\\
&&\\
Z^{(3)}_{0,m}&=&\frac{1}{\sqrt{2}}\left(\sqrt{2}Z^{(1)}_{0,m}-Z_{-1,m-1}^{(1)}+Z_{0,m-1}^{(3)}\right),\nonumber\\
&&
\end{eqnarray}
\end{subequations}
while for $m<0$ Eq.~(\ref{bulk}) still holds.
The solution
$(Z^{(1)}_{m,n},Z^{(3)}_{m,n})\propto(1,\sqrt{2}-1)$ obeys
the infinite mass boundary condition (\ref{infmass2}) with $\alpha=-\pi/4$, as
required.

This completes the boundary conditions for the four orientations $a$, $b$, $c$, and $d$.
The orientations $a'$, $b'$, $c'$, and $d'$ are obtained by the following symmetry:
The network model is left invariant by a $\pi$ rotation
in coordinate space (which takes $\bm{r}$ to $-\bm{r}$) together with the
application of $\sigma_y$ in spinor space (which takes $Z^{(1)}$ to $-iZ^{(3)}$ and $Z^{(3)}$ to $iZ^{(1)}$).

\section{Stable method of multiplication of transfer matrices}
\label{numapp}
To construct the transfer matrix of a conductor one can divide it into slices, compute the transfer matrix of each slice, and multiply the individual transfer matrices. This recursive construction is numerically unstable, because products of transfer matrices contain exponentially growing eigenvalues which overwhelm the small eigenvalues relevant for transport properties. Chalker and Coddington \cite{Cha88} used an orthogonalisation method \cite{Pic81,Mac83} to calculate the small eigenvalues in a numerically stable way. To obtain both eigenvalues and eigenfunctions we employ an alternative method \cite{Tam91,Bar07}: Using the condition of current conservation, the product of transfer matrices can be converted into a composition of unitary matrices, involving only eigenvalues of unit absolute value.

We briefly outline how the method works for the real space transfer matrices $Y$ of the network model, defined by Eq.\ \eqref{defy}. For the recursive construction it is convenient to rewrite this definition as
\begin{equation}
\left(\begin{array}{r}Z^{(1)}_{m+L,m-L}\\Z^{(3)}_{m+L,m-L}\end{array}\right)=\sum_{n=0}^{N-1}Y(L,L')_{m,n}
\left(\begin{array}{r}Z^{(1)}_{n+L',n-L'}\\Z^{(3)}_{n+L',n-L'}\end{array}\right).
\end{equation}
The numbers $L,L'$ are integers, so that $Y(L,L')$ is the transfer matrix from $x'=2L'l$ to $x=2Ll$. The composition law for transfer matrices is matrix multiplication,
\begin{equation}
Y(L,0)=Y(L,L-1)Y(L-1,0),\label{recurst}
\end{equation}
with initial condition $Y(0,0)=$ identity matrix. 

The unstable matrix multiplication may be stabilized with the help of the condition $Y^{-1}=\Sigma_{z}Y^{\dagger}\Sigma_{z}$ of current conservation (see Sec.\ \ref{sect4}). Because of this condition, the matrix $U$ constructed from $Y$ by
\begin{equation}
Y=\begin{pmatrix}
a&b\\
c&d
\end{pmatrix}
\Leftrightarrow
U=\begin{pmatrix}
-d^{-1}c&d^{-1}\\
a-bd^{-1}c&bd^{-1}
\end{pmatrix}\label{YtoU}
\end{equation}
is a unitary matrix ($U^{-1}=U^{\dagger}$). Matrix multiplication of $Y$'s induces a nonlinear composition of $U$'s,
\begin{equation}
Y_{1}Y_{2}\Leftrightarrow U_{1}\otimes U_{2},
\end{equation}
defined by
\begin{equation}
\begin{pmatrix}
a_1&b_1\\
c_{1}&d_{1}
\end{pmatrix}
\otimes
\begin{pmatrix}
a_2&b_2\\
c_{2}&d_{2}
\end{pmatrix}
=\begin{pmatrix}
a_3&b_3\\
c_3&d_3
\end{pmatrix},
\end{equation}
\begin{subequations}
\begin{eqnarray}
a_3&=&a_1+b_1(1-a_2d_1)^{-1}a_2c_1,\\
b_3&=&b_1(1-a_2d_1)^{-1}b_2,\\
c_3&=&c_2(1-d_1a_2)^{-1}c_1,\\
d_3&=&d_2+c_2(1-d_1a_2)^{-1}d_1b_2.
\end{eqnarray}
\end{subequations}

The algorithm now works as follows: Multiply a number of transfer matrices and stop well before numerical overflow would occur. Transform this transfer matrix into a unitary matrix according to Eq.\ (\ref{YtoU}). Continue with the next sequence of transfer matrices, convert to a unitary matrix and convolute with the previous unitary matrix. At the end, we may transform back from $U$ to $Y$ by the inverse of relation \eqref{YtoU}
\begin{equation}
U=\begin{pmatrix}
A&B\\
C&D
\end{pmatrix}
\Leftrightarrow
Y=\begin{pmatrix}
C-DB^{-1}A&DB^{-1}\\
-B^{-1}A&B^{-1}
\end{pmatrix}.\label{UtoY}
\end{equation}
In practice this final transformation is unnecessary. According to Eq.~(\ref{YtoU}) the upper-right block of $U$ is $d^{-1}\equiv(Y^{--})^{-1}$,
which is all we need to calculate the conductance using the Landauer formula (\ref{gnet}).  
 
\section{Optimal choice of phases in the network model}
\label{phase}
In Sec.~\ref{five} we noted that the same long-wavelength correspondence between
the Dirac equation and the network model can be obtained for
different choices of the phases $\phi^{(k)}_{m,n}$. Among these choices,
the choice (\ref{choice2}) avoids corrections of order $\bm{\partial_r}V\,l$
to the Dirac equation. Here we show why.

For $\mu=\bm{A}=0$ Eq.~(\ref{corresp}) reduces to
\begin{subequations}
\begin{align}
&\beta_{m,n}=0,\\
&\phi^{(1)}_{m,n}=\phi^{(3)}_{m,n}=(1-\alpha)\varepsilon_{m,n},\\
&\phi^{(2)}_{m,n}=\phi^{(4)}_{m,n}=\alpha\varepsilon_{m,n},
\end{align}
\end{subequations}
where we have defined the dimensionless quantity $\varepsilon_{m,n}\equiv\left[E-V(\bm{r}_{m,n})\right]l/\hbar v$.
The parameter $\alpha$ can be chosen arbitrarily. We wish to show that the choice $\alpha=0$ is optimal.
We substitute Eq.~(\ref{neteq1}) into Eq.~(\ref{neteq2}) of Sec.~\ref{sect3}, with this parametrization, and obtain
\begin{widetext}
\begin{subequations}
\label{fulleq}
\begin{align}
Z_{m,n}^{(1)}=&\frac{e^{i\varepsilon_{m,n}}}{2}\left[e^{-i\alpha (\varepsilon_{m,n+1}-\varepsilon_{m,n})}(Z_{m,n+1}^{(1)}+Z^{(3)}_{m+1,n+1})
+Z_{m-1,n}^{(1)}-Z_{m,n}^{(3)}\right],\\
Z_{m,n}^{(3)}=&\frac{e^{i\varepsilon_{m,n}}}{2}\left[Z_{m,n}^{(1)}+Z_{m+1,n}^{(3)}
-e^{-i\alpha (\varepsilon_{m,n-1}-\varepsilon_{m,n})}(Z_{m-1,n-1}^{(1)}-Z^{(3)}_{m,n-1})\right].
\end{align}
\end{subequations}
\end{widetext}
Now we expand in $\varepsilon_{m,n}$, keeping terms to first order, and take $Z^{(1)}$ and $Z^{(3)}$
to be functions defined for all $\bm{r}$ and smooth on the scale of the lattice.
From Eq.~(\ref{fulleq}) we then obtain
\begin{align}
&0=\left[E+\sigma_z p_x+\sigma_x p_y-V(\bm{r})\right]
\left(\begin{array}{r}Z^{(1)}\\Z^{(3)}\end{array}\right)\nonumber\\
&-\frac{\alpha}{2}\left(\begin{array}{rr}
V(\bm{r}+\bm{a}_2)-V(\bm{r})&V(\bm{r}+\bm{a}_2)-V(\bm{r})\\
V(\bm{r})-V(\bm{r}-\bm{a}_2)&V(\bm{r}-\bm{a}_2)-V(\bm{r})\end{array}\right)
\left(\begin{array}{r}Z^{(1)}\\Z^{(3)}\end{array}\right).\label{expand}
\end{align}
After transforming to $\Psi={\cal G}(Z^{(1)},Z^{(3)})^{T}$, with $\cal G$ as in Eq.~(\ref{connection}),
the first term on the r.h.s. of Eq.~(\ref{expand}) becomes the desired Dirac equation. If we choose $\alpha\not=0$
then the potential $V$ has to be smooth on the scale of the lattice, 
for the second term to be negligible in comparison with the first. We conclude that $\alpha=0$ is the optimal choice.

\end{document}